\documentclass[twocolumn,superscriptaddress,amsmath,amssymb,aps,pra]{revtex4-2}
\usepackage{graphicx}
\usepackage{CJKutf8}
\usepackage{dcolumn}
\usepackage{amsmath,bm}
\usepackage[mathlines]{lineno}
\usepackage{color}
\usepackage[colorlinks=true,linkcolor=blue,citecolor=magenta,urlcolor=cyan]{hyperref}
\usepackage{orcidlink}

\begin{document}
\title{Band inversion transition in HgTe nanowire grown along the [001] direction}
\author{Rui\! Li~(\begin{CJK}{UTF8}{gbsn}李睿\end{CJK})\,\orcidlink{0000-0003-1900-5998}}
\email{ruili@ysu.edu.cn}
\affiliation{Hebei Key Laboratory of Microstructural Material Physics, School of Science, Yanshan University, Qinhuangdao 066004, China}

\begin{abstract}
{\color{red}}
    The low-energy effective Hamiltonian of a cylindrical HgTe nanowire grown along the [001] crystallographic direction is constructed by using the perturbation theory. Both the anisotropic term  and the bulk inversion asymmetry term of the Kane model are taken into account. Although the anisotropic term has converted the crossing between the $E_{1}$ and $H_{1}$ subbands into an anticrossing at $k_{z}R\!=\!0$, the gap-closing-and-reopening transition in the subband structure can still occur at finite wave vectors $k_{z}R\!\approx\!\pm0.24$ for critical nanowire radius $R\!\approx\!3.45$ nm. The bulk inversion asymmetry does not contribute to the low-energy effective Hamiltonian, i.e., there is no spin splitting in the $E_{1}$, $H_{1}$, and $H_{2}$ subbands for a [001] oriented cylindrical nanowire. 
  \end{abstract}
%\date{November 8, 2025}
\date{\today}
\maketitle

\section{Introduction}
{\color{red}}
%topological insulator
The gap-closing-and-reopening transition in the band structure of a given material or quantum structure is a signature of the topological phase transition~\cite{doi:10.1126/science.1133734,RevModPhys.83.1057,RevModPhys.82.3045}. This kind of phase transition is characterized by a change of the topological invariant~\cite{PhysRevLett.62.2747,PhysRevLett.49.405,PhysRevLett.95.146802}. One of the gapped phases must be topologically nontrivial, and the material or quantum structure is said to be a topological insulator~\cite{bernevig2013topological,shen2012topological} when it is in this phase. The Su-Schrieffer-Heeger model~\cite{PhysRevLett.42.1698} of polyacetylene is one example, where the topological invariant is the Zak phase~\cite{PhysRevLett.62.2747}. The Haldane model of graphene is another example~\cite{PhysRevLett.61.2015}, where the topological invariant is the Chern number~\cite{PhysRevLett.49.405}. The Kane-Mele model of graphene proposed in 2005 is also topological~\cite{PhysRevLett.95.226801}, the topological invariant in this case is the $Z_{2}$ number~\cite{PhysRevLett.95.146802}. 

%HgTe nanowire
Semiconductor quantum structures, e.g., quantum wells~\cite{doi:10.1126/science.1148047,PhysRevB.72.035321,PhysRevB.105.035305} and nanowires~\cite{Selvig_2006,ridene2018mid,RL_2025b}, are ideal platforms for searching the topological insulator phase, because the band structure of these structures is engineerable. Indeed, topological insulator phase has been predicted and verified in HgTe/CdTe~\cite{doi:10.1126/science.1133734,doi:10.1126/science.1148047,doi:10.1143/JPSJ.77.031007,PhysRevLett.101.246807} and InAs/GaSb/AlSb~\cite{PhysRevLett.100.236601,PhysRevLett.107.136603,PhysRevLett.114.096802} quantum wells, where the gap-closing-and-reopening transition is induced by tuning the quantum well width. Inspired by the above success in quasi-two-dimensional systems, the topological insulator phase has also been searched in quasi-one-dimensional HgTe~\cite{RL_2025a,RL2025c}, SnTe~\cite{doi:10.1021/acsanm.4c00506}, and InAs/GaSb~\cite{1pwr-lzz5} nanowires. In particular, based on the Kane model in the isotropic approximation, the gap-closing-and-reopening transition in the HgTe nanowire is demonstrated~\cite{RL_2025a,RL2025c}. The negative band gap of bulk HgTe is essential for the occurrence of the gap-closing-and-reopening transition in both quasi-one and two-dimensions. Note that several experiments have reported the successful growth of quasi-one-dimensional HgTe nanowires~\cite{Selvig_2006,PhysRevMaterials.1.023401}.  

%this paper
Here we want to address two issues that are overlooked in the previous studies of the quasi-one-dimensional HgTe nanowire~\cite{RL_2025a,RL2025c}, i.e., the effects of both the anisotropic term and the bulk inversion asymmetry term~\cite{PhysRev.100.580,winkler2003spin} in the Kane model~\cite{KANE1957249}. We construct the low-energy effective Hamiltonian in the HgTe nanowire by incorporating the above two effects. The effective Hamiltonian does not have the Dirac form~\cite{doi:10.1098/rspa.1928.0023}, and is actually a six-band model with  two separate $3\!\times\!3$ blocks. The anisotropic term in the Kane model forces the crossing between the $E_{1}$ and $H_{1}$ subbands to become an anticrossing at $k_{z}R\!=\!0$. There still exists a gap-closing-and-reopening transition in the subband structure when we tune the nanowire radius. However, the transition point has been forced to a place with wave vectors $k_{z}R\!\neq\!0$. The bulk inversion asymmetry does not contribute to the low-energy effective Hamiltonian, as all the matrix elements of the inversion asymmetry term $H_{\rm BIA}$ vanish in the low-energy Hilbert subspace. For radius larger than $R\!\approx\!3.45$ nm, the [001] oriented HgTe nanowire is in the topological insulator regime, where end states can be found in the presence of open boundary condition.

\section{Description of the model}
{\color{red}}
The negative band gap of bulk HgTe plays an essential role in the band inversion transition of a HgTe/CdTe quantum well~\cite{doi:10.1126/science.1133734}. The negative band gap refers to the lying of the $\Gamma_{6}$ band below the $\Gamma_{8}$ band~\cite{madelung2004semiconductors}. It follows that the minimal bulk model describing the band structure near the $\Gamma$ point of HgTe should be a six-band Kane model~\cite{KANE1957249}. Also, previous studies have revealed that the contribution from the spin-orbit split-off $\Gamma_{7}$ band to the low-energy subband states is negligible~\cite{doi:10.1126/science.1133734,Pfeuffer2000}. Here, we also use the six-band Kane model in studying a cylindrical HgTe nanowire. When the wave vectors are along the crystal cubic axes, i.e., $k_{x}\!\parallel\![100]$, $k_{y}\!\parallel\![010]$, and $k_{z}\!\parallel\![001]$, the six-band Kane model can be conveniently decomposed into three parts~\cite{winkler2003spin}
\begin{equation}
    H_{\rm Kane}=H_{0}+H'+H'',\label{eq_Kanemodel}
\end{equation}
where
\begin{widetext}
\begin{eqnarray}
    H_{0}&=&\frac{\hbar^{2}}{2m_{0}}\left(\begin{array}{cccccc}
        \varepsilon_{0}+k^{2}_{\parallel}&0&-\frac{1}{\sqrt{2}}Pk_{+}&0&\frac{1}{\sqrt{6}}Pk_{-}&0\\
        0&\varepsilon_{0}+k^{2}_{\parallel}&0&-\frac{1}{\sqrt{6}}Pk_{+}&0&\frac{1}{\sqrt{2}}Pk_{-}\\
        -\frac{1}{\sqrt{2}}Pk_{-}&0&-(\gamma_{1}+\gamma_{2})k^{2}_{\parallel}&0&\sqrt{3}\gamma_{2}k^{2}_{-}&0\\
        0&-\frac{1}{\sqrt{6}}Pk_{-}&0&-(\gamma_{1}-\gamma_{2})k^{2}_{\parallel}&0&\sqrt{3}\gamma_{2}k^{2}_{-}\\
        \frac{1}{\sqrt{6}}Pk_{+}&0&\sqrt{3}\gamma_{2}k^{2}_{+}&0&-(\gamma_{1}-\gamma_{2})k^{2}_{\parallel}&0\\
        0&\frac{1}{\sqrt{2}}Pk_{+}&0&\sqrt{3}\gamma_{2}k^{2}_{+}&0&-(\gamma_{1}+\gamma_{2})k^{2}_{\parallel}\end{array}\right),\nonumber\\
        H'&=&\frac{\hbar^{2}}{2m_{0}}\left(\begin{array}{cccccc}k^{2}_{z}&0&0&\sqrt{\frac{2}{3}}Pk_{z}&0&0\\
        0&k^{2}_{z}&0&0&\sqrt{\frac{2}{3}}Pk_{z}&0\\
        0&0&-(\gamma_{1}-2\gamma_{2})k^{2}_{z}&2\sqrt{3}\gamma_{3}k_{z}k_{-}&0&0\\
        \sqrt{\frac{2}{3}}Pk_{z}&0&2\sqrt{3}\gamma_{3}k_{z}k_{+}&-(\gamma_{1}+2\gamma_{2})k^{2}_{z}&0&0\\
        0&\sqrt{\frac{2}{3}}Pk_{z}&0&0&-(\gamma_{1}+2\gamma_{2})k^{2}_{z}&-2\sqrt{3}\gamma_{3}k_{z}k_{-}\\
        0&0&0&0&-2\sqrt{3}\gamma_{3}k_{z}k_{+}&-(\gamma_{1}-2\gamma_{2})k^{2}_{z}
        \end{array}\right), \nonumber
\end{eqnarray}
\end{widetext}
and 
\begin{equation}
    H''=\frac{\sqrt{3}\hbar^{2}}{4m_{0}}(\gamma_{3}-\gamma_{2})(k^{2}_{+}-k^{2}_{-})\left(\begin{array}{cccccc}0&0&0&0&0&0\\0&0&0&0&0&0\\0&0&0&0&-1&0\\0&0&0&0&0&-1\\0&0&1&0&0&0\\0&0&0&1&0&0\end{array}\right),
\end{equation}
with $m_{0}$ being the free electron mass, $k^{2}_{\parallel}\!=\!k^{2}_{x}+k^{2}_{y}$, $k_{\pm}\!=\!k_{x}\pm\,ik_{y}$, and $\hbar^{2}\varepsilon_{0}/(2m_{0})\!=\!-0.303$ eV, $\hbar^{2}P^{2}/(2m_{0})\!=\!18.8$ eV, $\gamma_{1}\!=\!4.1$, $\gamma_{2}\!=\!0.5$, and $\gamma_{3}\!=\!1.3$ being the bulk band parameters of HgTe~\cite{PhysRevB.72.035321}.

In Eq.~(\ref{eq_Kanemodel}), $H_{0}$, $H'$, and $H''$ are the isotropic term, the $k_{z}\!\neq\!0$ term, and the anisotropic term of the Kane model, respectively. We have deliberately decomposed the Kane model into three parts listed above for the purpose that we want to regard both $k_{z}$ and $\gamma_{3}-\gamma_{2}$ as small quantities, such that both $H'$ and $H''$ will be treated as perturbations in the following calculations. 

We now introduce the nanowire model we are considering. A cylindrical HgTe nanowire grown along the [001] crystallographic direction is under investigation. In the framework of the envelope function approximation~\cite{winkler2003spin}, the size quantization in the nanowire is governed by the following multiband Hamiltonian~\cite{ridene2018mid,RL_2025a}
\begin{equation}
    H=H_{\rm Kane}+V(r),\label{eq_multibandH}
\end{equation}
where $V(r)$ is the hard-wall confining potential
\begin{equation}
    V(r)=\left\{\begin{array}{cc}0,&r<R,\\\pm\infty,&r>R,\end{array}\right.
\end{equation}
with $R$ being the radius of the nanowire. Here the $+/-$ signs apply to the $\Gamma_{6/8}$ bands respectively, and we have also chosen to study in a cylindrical representation where $x\!=\!r\cos\varphi$, $y\!=\!r\sin\varphi$, and $z\!=\!z$, such that $r\!=\!\sqrt{x^{2}+y^{2}}$. Note that the replacement $k_{x,y}\!=\!-i\partial_{x,y}$ should be made in the multiband Hamiltonian (\ref{eq_multibandH}), and $k_{z}$ is a conserved quantity. Note that currently we do not include the bulk inversion asymmetry term in Hamiltonian~(\ref{eq_multibandH}), the effect of this term will be addressed in Sec.\ref{sec_BIA}.

\section{Low-energy effective Hamiltonian}
{\color{red}}
In order to obtain the low-energy effective Hamiltonian describing the subband structure of the nanowire close to the wave vector $k_{z}=0$, we first solve the zeroth order Schr\"odinger equation $[H_{0}+V(r)]\Psi(r,\varphi)\!=\!E_{0}\Psi(r,\varphi)$. Because $H_{0}$ is isotropic, the zeroth order Schr\"odinger equation can be solved exactly using the method of Sercel and Vahala~\cite{PhysRevB.42.3690}. The lowest electron subband states $|E_{1},\pm1/2\rangle$ and the two top most hole subband states $|H_{1,2},\pm1/2\rangle$ are obtained and used as the basis states for constructing the effective Hamiltonian (see also Ref.~\cite{RL2025c}). We then consider both $H'$ and $H''$ in Eq.~(\ref{eq_multibandH}) as perturbations, and calculate their matrix elements in the Hilbert subspace spanned by the ordered set $|E_{1},1/2\rangle$, $|H_{1},1/2\rangle$, $|H_{2},1/2\rangle$, $|E_{1},-1/2\rangle$, $|H_{1},-1/2\rangle$, and $|H_{2},-1/2\rangle$.

\begin{table*}
\caption{\label{tab1}Parameters of the effective Hamiltonian $H_{+}(k_{z})$.}
\begin{ruledtabular}
\begin{tabular}{cccccccccccccc}
$R$~(nm)&$A$~(eV)&$B$~(eV)&$C$~(eV)&$D$~(eV)&$F$~(eV)&$G$~(eV)&$\Delta_{e}$~(eV)&$\Delta_{h1}$~(eV)&$\Delta_{h2}$~(eV)&$\zeta_{e}$~(eV)&$\zeta_{h_{1}}$~(eV)&$\zeta_{h_{2}}$~(eV)&$\zeta_{eh_{1}}$~(eV)\\
3.10&0.0024&-0.0182&-0.0140&0.0008&0.1222&0.0727&-0.0927&-0.1804&-0.2394&-0.0116&0.0349&0.0167&0.0049\\
3.45&0.0019&-0.0147&-0.0115&0.0007&0.0921&0.0601&-0.1116&-0.1457&-0.2105&-0.0107&0.0282&0.0200&0.0039\\
3.80&0.0015&-0.0121&-0.0098&0.0006&0.0655&0.0482&-0.1269&-0.1201&-0.1855&-0.0100&0.0232&0.0228&0.0032
\end{tabular}
\end{ruledtabular}
\end{table*}

In this way, we are able to obtain the following $6\times6$ effective Hamiltonian 
\begin{equation}
H_{\rm ef}(k_{z})=\left(\begin{array}{cc}H_{+}(k_{z})&0\\0&H^{*}_{+}(-k_{z})\end{array}\right),\label{eq_effHamiltonian}
\end{equation}
where
\begin{eqnarray}
H_{+}(k_{z})&=&\left(\begin{array}{ccc}Ak^{2}_{z}+\Delta_{e}&-iDk^{2}_{z}&Fk_{z}\\
iDk^{2}_{z}&Bk^{2}_{z}+\Delta_{h1}&-iGk_{z}\\
Fk_{z}&iGk_{z}&Ck^{2}_{z}+\Delta_{h2}\end{array}\right)\nonumber\\
&&+\left(\begin{array}{ccc}\zeta_{e}&-i\zeta_{eh_{1}}&0\\i\zeta_{eh_{1}}&\zeta_{h_{1}}&0\\0&0&\zeta_{h_{2}}\end{array}\right),\label{eq_efHamil}
\end{eqnarray}
with $k_{z}$ in units of $1/R$ and $H^{*}_{+}(-k_{z})$ in the lower block being a consequence of time reversal symmetry~\cite{doi:10.1126/science.1133734}. Here $\Delta_{e}$ and $\Delta_{h_{1,2}}$ are the zeroth order energy eigenvalues for the lowest electron and the two top most hole subbands, respectively, the $k_{z}$ terms are given by the perturbation $H'$, and the $\zeta$ terms are given by the perturbation $H''$. Certainly, all of the parameters, i.e., $\Delta_{e,h_{1,2}}$, $A\ldots\,G$, and $\zeta_{e,h_{1,2},eh_{1}}$, are implicit functions of the nanowire radius $R$~\cite{RL2025c}. Some representative values of these parameters are given in Tab.~\ref{tab1} for selected radii $R=3.10$ nm, $3.45$ nm, and $3.80$ nm. Note that $k_{z}$ is treated as a perturbation parameter in our derivations, such that the effective Hamiltonian (\ref{eq_effHamiltonian}) is valid only for small $k_{z}$'s, i.e., $|k_{z}R|\!\ll\!1$.

\begin{figure}
\includegraphics{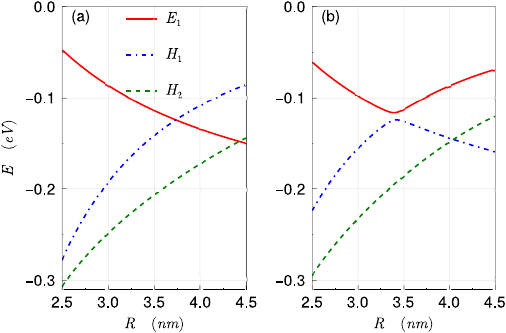}
\caption{\label{fig_evsr}The lowest electron and the two top most hole subband energies at $k_{z}=0$ as functions of the nanowire radius $R$. (a) Result without the anisotropic term $H''$. (b) Result with the anisotropic term $H''$. The anisotropic term $H''$ causes the crossing between the $E_{1}$ and $H_{1}$ subbands to become an anticrossing. The anticrossing occurs at $R\approx3.4$ nm, and the energy splitting at the anticrossing is about $8.2$ meV.} 
\end{figure}

We first focus on the impact of the anisotropic term $H''$ on the subband energies at $k_{z}\!=\!0$. We directly diagonalize $H_{\rm ef}(k_{z}\!\!\rightarrow\!0)$ to obtain the subband energies at $k_{z}\!=\!0$, and plot them as functions of the nanowire radius $R$. Without the anisotropic term $H''$, the lowest electron subband $E_{1}$ crosses with the two top most hole subbands $H_{1}$ and $H_{2}$ [see Fig.~\ref{fig_evsr}(a)]. This phenomenon has already been predicted in our previous paper~\cite{RL_2025a}. Note that the crossing positions obtained here are slightly different from that obtained previously. This is because the isotropic approximation~\cite{PhysRev.102.1030,PhysRevB.8.2697} used here $\gamma_{s}=\gamma_{2}$ is different from that used previously $\gamma_{s}=(2\gamma_{2}+3\gamma_{3})/5$. When the anisotropic term $H''$ is included, the crossing between the $E_{1}$ and $H_{1}$ subbands becomes an anticrossing [see Fig.~\ref{fig_evsr}(b)], while the second crossing involving the $H_{2}$ subband is still a crossing although its position has changed a little bit. 

\begin{figure*}
\includegraphics{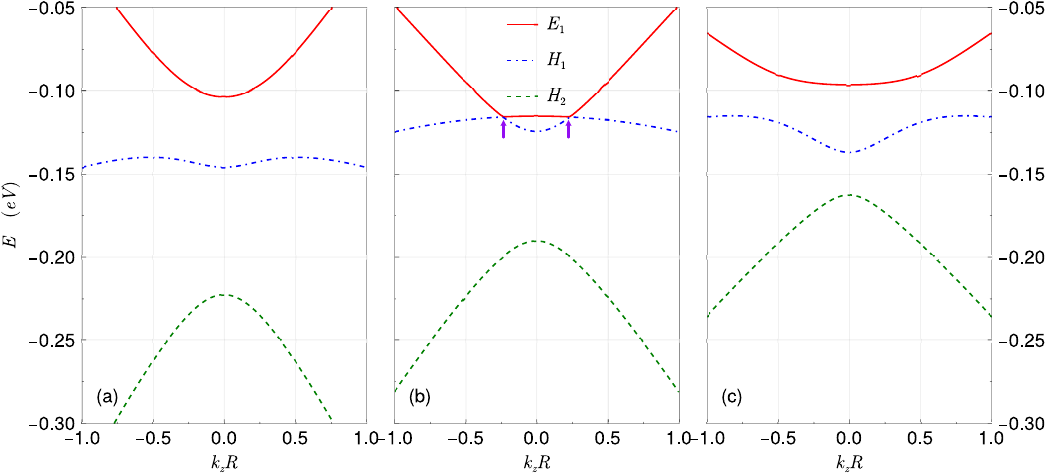}
\caption{\label{fig_effband}The lowest electron subband dispersion $E_{1}$ and the two top most hole subband dispersions $H_{1}$ and $H_{2}$ near $k_{z}R=0$. (a) Result for radius $R=3.10$ nm with normal band structure. (b) Result for radius $R=3.45$ nm with critical band structure, the subband gap closes at $k_{z}R\approx\pm0.24$. The arrows mark the points where the gap-closing-and-reopening occurs. (c) Result for radius $R=3.80$ nm with inverted band structure.}
\end{figure*}

We next examine the impact of the anisotropic term $H''$ on the subband energies at $k_{z}\!\neq\!0$. Since the anisotropic term has prevented the gap-closing-and-reopening transition from occurring at the wave vector $k_{z}\!=\!0$, it is an interesting question whether this transition still exists when we tune the nanowire radius $R$. Via diagonalizing the effective Hamiltonian (\ref{eq_effHamiltonian}) for finite $k_{z}$'s, we find the position of the gap-closing-and-reopening transition has moved to the wave vectors $k_{z}\!\neq\!0$. In Figs.~\ref{fig_effband}(a), (b), and (c), we show the subband dispersions for HgTe nanowires with radii $R\!=\!3.10$ nm, $3.45$ nm, and $3.80$ nm, respectively. The subband gap closes at $k_{z}R\approx\pm0.24$ for critical nanowire radius $R\!\approx\!3.45$ nm. Note that the band parameters for nanowire radii $R\!=\!3.10$ nm, $3.45$ nm, and $3.80$ nm are given in Tab.~\ref{tab1}. We emphasize that the position of the transition point $k_{z}R\!\approx\!\pm0.24$ roughly lies in the valid range of the effective Hamiltonian $|k_{z}R|\!\ll\!1$.

Similar phenomenons have also been observed in the HgTe quantum well, where the bulk inversion asymmetry plays the role of the anisotropic term here~\cite{PhysRevB.77.125319,WINKLER20122096,PhysRevB.91.081302,Durnev:2022aa}. The bulk inversion asymmetry causes an anticrossing between the $E_{1}$ and $H_{1}$ subbands at $k_{\parallel}=0$ in the quantum well, and the position of the gap-closing-and-reopening transition has moved to the nonzero in-plane wave vectors $k_{\parallel}$~\cite{PhysRevB.77.125319,WINKLER20122096,PhysRevB.91.081302,Durnev:2022aa}. We note that the effective Hamiltonian in the quantum well is of the Dirac form that contains four bands~\cite{doi:10.1098/rspa.1928.0023}, while the effective Hamiltonian in our case contains six bands [see Eq.~(\ref{eq_effHamiltonian})]. We also note that the band structure shown in Fig.~\ref{fig_effband}(b) is very similar to that of a compressively strained bulk HgTe or $\alpha$-Sn~\cite{PhysRevB.95.201101,PhysRevB.97.195139}, although their underlying physics are different.

\begin{figure}
\includegraphics{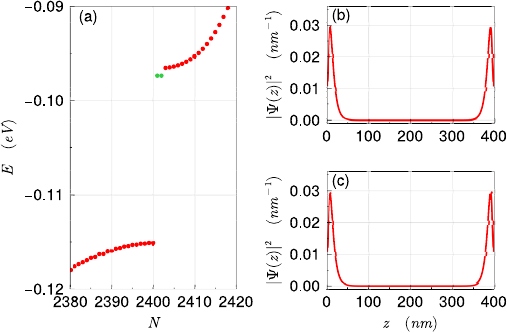}
\caption{\label{fig_endstate}The energies in the finite nanowire with radius $R=3.8$ nm and length 400 nm versus the state number $N$ (a). There are two energy states (marked green) in the subband gap. The density probability distributions of  these two energy states are given in (b) and (c), respectively. Note that only the upper block of the effective Hamiltonian $H_{+}(k_{z})$ is considered here. }
\end{figure}

Since there is still a gap-closing-and-reopening transition in the subband structure via tuning the nanowire radius, one of the gapped phases must be topologically nontrivial. Based on our previous results using the isotropic approximation, we deduce the nanowire is actually a quasi-one-dimensional topological insulator for radius $R\!>\!3.45$ nm~\cite{RL2025c}. Indeed, we are able to obtain the end states via numerically solving the effective Hamiltonian $H_{+}(k_{z})$ with open boundary condition. The energy spectrum for a finite nanowire with radius $R\!=\!3.80$ nm and length 400 nm is shown in Fig.~\ref{fig_endstate}(a). There exist two energy states that have been marked green in the subband gap. The density probability distributions of these two states are shown in Fig.~\ref{fig_endstate}(b) and (c), respectively. These two states can be regarded as superposition states of the left and the right end states, and there is no obvious difference between Figs.~\ref{fig_endstate}(b) and (c). Note that the results obtained here are very similar to that obtained using the isotropic approximation~\cite{RL2025c}, and the topological insulator phase demonstrated here is similar to that in strained bulk HgTe or $\alpha$-Sn~\cite{PhysRevB.77.125319,PhysRevB.76.045302,PhysRevLett.106.126803,PhysRevB.83.075110,PhysRevLett.111.157205}.

In one-dimensional systems, the Zak phase is usually used as a topological invariant~\cite{PhysRevLett.62.2747}. For our $3\times3$ effective model (\ref{eq_efHamil}), it is hard to obtain its analytical eigenstates, such that currently we are unable to calculate the corresponding topological invariant. However, the existence of end states is another signature of the topologically nontrivial phase. The topological sense in our model can be explained as follows. When the parameters of the effective model (\ref{eq_efHamil}) vary continuously, e.g., we tune the nanowire radius $R$, as long as the subband gap does not close, the end states always present in the nontrivial region ($R\!>\!3.45$ nm), while there is always no end state in the trivial region ($R\!<\!3.45$ nm). Therefore, the presence or absence of the end states has topological origin.

\section{Transformation of representation}
{\color{red}}
We can see from Eq.~(\ref{eq_efHamil}) that the effective Hamiltonian is not diagonal at $k_{z}\!=\!0$. The off-diagonal elements at $k_{z}\!=\!0$ are essentially induced by the anisotropic term $H''$ of the Kane model (\ref{eq_Kanemodel}). In order to explicitly know the subband energies at $k_{z}\!=\!0$, here we perform a representation transformation such that the effective Hamiltonian is diagonal at $k_{z}\!=\!0$.

We first focus on the $k_{z\!}=\!0$ part of the effective Hamiltonian $H_{+}(k_{z})$
\begin{equation}
    H_{+}(k_{z}\!\rightarrow\!0)=\left(\begin{array}{ccc}\Delta_{e}+\zeta_{e}&-i\zeta_{eh_{1}}&0\\i\zeta_{eh_{1}}&\Delta_{h_{1}}+\zeta_{h_{1}}&0\\0&0&\Delta_{h_{2}}+\zeta_{h_{2}}\end{array}\right).
\end{equation}
Although this Hamiltonian is already in a block diagonal form, we can further diagonalize it using the following unitary matrix
\begin{equation}
    U=\left(\begin{array}{ccc}\cos\frac{\theta}{2}&\sin\frac{\theta}{2}&0\\i\sin\frac{\theta}{2}&-i\cos\frac{\theta}{2}&0\\0&0&1\end{array}\right),
\end{equation}
where 
\begin{equation}
    \theta=\pi+\arctan\frac{2\zeta_{eh_{1}}}{\Delta_{e}+\zeta_{e}-\Delta_{h_{1}}-\zeta_{h_{1}}}.
%    \cos\theta&=&\frac{\frac{1}{2}\left(\Delta_{e}+\zeta_{e}-\Delta_{h_{1}}-\zeta_{h_{1}}\right)}{\sqrt{\frac{1}{4}\left(\Delta_{e}+\zeta_{e}-\Delta_{h_{1}}-\zeta_{h_{1}}\right)^{2}+\zeta^{2}_{eh_{1}}}},\nonumber\\
%    \sin\theta&=&\frac{\zeta_{eh_{1}}}{\sqrt{\frac{1}{4}\left(\Delta_{e}+\zeta_{e}-\Delta_{h_{1}}-\zeta_{h_{1}}\right)^{2}+\zeta^{2}_{eh_{1}}}}.
\end{equation}
Note that the period $\pi$ should been added in the above equation if the denominator in the arctan function is negative. One can verify the unitary property $U^{\dagger}U\!=\!UU^{\dagger}\!=\!1$. After a unitary transformation, we have the diagonal Hamiltonian at $k_{z}\!=\!0$
\begin{equation}
    U^{\dagger}H_{+}(k_{z}\!\rightarrow\!0)U=\left(\begin{array}{ccc}\Delta'_{e}&0&0\\0&\Delta'_{h_{1}}&0\\0&0&\Delta'_{h_{2}}\end{array}\right),
\end{equation}
where 
\begin{eqnarray}
    \Delta'_{e}&=&\frac{1}{2}\left(\Delta_{e}+\zeta_{e}+\Delta_{h_{1}}+\zeta_{h_{1}}\right)\nonumber\\
    &&+\sqrt{\frac{1}{4}\left(\Delta_{e}+\zeta_{e}-\Delta_{h_{1}}-\zeta_{h_{1}}\right)^{2}+\zeta^{2}_{eh_{1}}},\nonumber\\
    \Delta'_{h_{1}}&=&\frac{1}{2}\left(\Delta_{e}+\zeta_{e}+\Delta_{h_{1}}+\zeta_{h_{1}}\right)\nonumber\\
    &&-\sqrt{\frac{1}{4}\left(\Delta_{e}+\zeta_{e}-\Delta_{h_{1}}-\zeta_{h_{1}}\right)^{2}+\zeta^{2}_{eh_{1}}},\nonumber\\
    \Delta'_{h_{2}}&=&\Delta_{h_{2}}+\zeta_{h_{2}}.
\end{eqnarray}
We find $\Delta'_{e}$ is always larger than $\Delta'_{h_{}1}$, regardless of the detailed values of $\Delta_{e,h_{1}}$ and $\zeta_{e,h_{1},eh_{1}}$. This result has actually been manifested in Fig.~\ref{fig_evsr}(b), where the $E_{1}$ energies always lie above the $H_{1}$ energies at $k_{z}\!=\!0$ when we tune the nanowire radius $R$. Hence, the anticrossing between the $E_{1}$ and $H_{1}$ subbands at $k_{z}\!=\!0$ is essentially caused by the anisotropic Hamiltonian $H''$.

\begin{table*}
\caption{\label{tab2}Parameters of the transformed effective Hamiltonian $U^{\dagger}H_{+}(k_{z})U$.}
\begin{ruledtabular}
\begin{tabular}{cccccccccc}
$R$~(nm)&$A'$~(eV)&$B'$~(eV)&$C'$~(eV)&$D'$~(eV)&$F'$~(eV)&$G'$~(eV)&$\Delta'_{e}$~(eV)&$\Delta'_{h1}$~(eV)&$\Delta'_{h2}$~(eV)\\
3.10&0.0023&-0.0181&-0.0140&0.0016&0.1129&0.0864&-0.1037&-0.1461&-0.2227\\
3.45&-0.0102&-0.0026&-0.0115&0.0074&-0.0075&0.1097&-0.1153&-0.1245&-0.1905\\
3.80&-0.0119&0.0013&-0.0098&0.0017&-0.0429&0.0691&-0.0966&-0.1372&-0.1627
\end{tabular}
\end{ruledtabular}
\end{table*}

We next apply the unitary transformation to the $k_{z}\!\neq\!0$ part of the effective Hamiltonian $H_{+}(k_{z})$. Combining the result of $k_{z}\!=\!0$ with that of the $k_{z}\!\neq\!0$, we have the following transformed effective Hamiltonian
\begin{equation}
    U^{\dagger}H_{+}U=\left(\begin{array}{ccc}A'k^{2}_{z}+\Delta'_{e}&D'k^{2}_{z}&F'k_{z}\\
D'^{*}k^{2}_{z}&B'k^{2}_{z}+\Delta'_{h1}&G'k_{z}\\
F'^{*}k_{z}&G'^{*}k_{z}&C'k^{2}_{z}+\Delta'_{h2}\end{array}\right),
\end{equation} 
where
\begin{eqnarray}
    A'&=&\frac{1}{2}\big(A+B+(A-B)\cos\theta+2 D \sin\theta\big),\nonumber\\
    B'&=&\frac{1}{2}\big(A+B-(A-B)\cos\theta-2D\sin\theta\big),\nonumber\\
    C'&=&C,\nonumber\\
    D'&=&\frac{1}{2}\big((A-B)\sin\theta-2D\cos\theta\big),\nonumber\\
    F'&=&F\cos\frac{\theta}{2}-G\sin\frac{\theta}{2},\nonumber\\
    G'&=&F\sin\frac{\theta}{2}+G\cos\frac{\theta}{2}.
\end{eqnarray}
We find that the transformed effective Hamiltonian has the same form as the first term of $H_{+}(k_{z})$ before the transformation [see Eq.~(\ref{eq_efHamil})]. Recall that the first term of $H_{+}(k_{z})$ is actually the effective Hamiltonian in the isotropic approximation~\cite{RL2025c}. In this respect, the effect of the anisotropic Hamiltonian $H''$ can be regarded as a renormalization to the parameters $A\ldots\,G$ and $\Delta_{e,h_{1,2}}$ of the effective Hamiltonian in the isotropic approximation. The renormalized parameters of the effective Hamiltonian near the gap closing-and-reopening transition are given in Tab.~\ref{tab2}. Note that when the nanowire radius increases, the sign of parameter $A’$ changes (from positive to negative) at about $R\!=\!3.35$ nm, while the sign of parameter $B’$ changes (from negative to positive) at about $R\!=\!3.50$ nm. The sign changes of $A’$ and $B’$ do not occur simultaneously.  This is possibly caused by the nonzero wave vector position ($k_{z}R\!\neq\!0$) of the gap-closing-and-reopening transition.

\section{\label{sec_BIA}The effect of the bulk inversion asymmetry}
{\color{red}}
The spin degeneracy in the band dispersions of a semiconductor is a consequence of both the time reversal symmetry and the space inversion symmetry~\cite{kittel1963quantum,winkler2003spin}. Bulk HgTe has a zinc blende lattice structure, in which there is no inversion center~\cite{madelung2004semiconductors}. It follows that there should exist spin splitting in the band structure due to this bulk inversion asymmetry. In the case of the six-band Kane model, the bulk inversion asymmetry is described by the following Hamiltonian~\cite{winkler2003spin}
\begin{widetext}
\begin{equation}
    H_{\rm BIA}=\left(\begin{array}{cccccc}
        0&0&{\rm h.c.}&{\rm h.c.}&{\rm h.c.}&{\rm h.c.}\\  
        0&0&{\rm h.c.}&{\rm h.c.}&{\rm h.c.}&{\rm h.c.}\\ 
        \frac{1}{\sqrt{2}}B_{+}k_{+}k_{z}&-\frac{1}{3\sqrt{2}}B_{-}(k^{2}_{x}+k^{2}_{y}-2k^{2}_{z})&0&{\rm h.c.}&{\rm h.c.}&{\rm h.c.}\\
        \frac{1}{\sqrt{6}}B_{-}(k^{2}_{x}-k^{2}_{y})-i\sqrt{\frac{2}{3}}B_{+}k_{x}k_{y}&\frac{1}{\sqrt{6}}B_{+}k_{+}k_{z}&-\frac{1}{2}\varkappa_{0}k_{-}&0&{\rm h.c.}&{\rm h.c.}\\
        \frac{1}{\sqrt{6}}B_{+}k_{-}k_{z}&-\frac{1}{\sqrt{6}}B_{-}(k^{2}_{x}-k^{2}_{y})-i\sqrt{\frac{2}{3}}B_{+}k_{x}k_{y}&\varkappa_{0}k_{z}&\frac{\sqrt{3}}{2}\varkappa_{0}k_{-}&0&{\rm h.c.}\\
        \frac{1}{3\sqrt{2}}B_{-}(k^{2}_{x}+k^{2}_{y}-2k^{2}_{z})&\frac{1}{\sqrt{2}}B_{+}k_{-}k_{z}&-\frac{\sqrt{3}}{2}\varkappa_{0}k_{+}&-\varkappa_{0}k_{z}&-\frac{1}{2}\varkappa_{0}k_{-}&0      
    \end{array}\right),
\end{equation}
\end{widetext}
where $B_{\pm}$ and $\varkappa_{0}$ are material parameters characterizing the strength of the inversion asymmetry and h.c. stands for Hermitian conjugation. Typical values of these parameters for HgTe can be found in Ref.~\cite{WINKLER20122096}. However, as we will show in the following, we do not need to know these values here. The bulk inversion asymmetry does not give rise to spin splitting in the [001] oriented cylindrical HgTe nanowire.

In the Hilbert subspace spanned by the ordered set $|E_{1},1/2\rangle$, $|H_{1},1/2\rangle$, $|H_{2},1/2\rangle$, $|E_{1},-1/2\rangle$, $|H_{1},-1/2\rangle$, and $|H_{2},-1/2\rangle$, we calculate the matrix elements of $H_{\rm BIA}$. Note that in the framework of the perturbation theory~\cite{landau1965quantum}, the $k_{z}$ terms in $H_{\rm BIA}$ are essentially second order small terms, such that we only need to calculate the matrix elements of $H_{\rm BIA}(k_{z}\!\rightarrow\!0)$. Detailed derivations show that all the matrix elements of $H_{\rm BIA}(k_{z}\!\rightarrow\!0)$ vanish, that means the bulk inversion asymmetry does not contribute to the effective Hamiltonian, and there is no spin splitting in the  [001] oriented cylindrical HgTe nanowire.

Our case is not the first in which the bulk inversion asymmetry does not give rise to spin splitting. In three dimension, the bulk inversion asymmetry does not give rise to spin splitting for both electrons and holes when ${\bf k}$ is parallel to [001]~\cite{winkler2003spin}. In quasi-two dimension, e.g., in a symmetric quantum well grown along the [110] direction, the bulk inversion asymmetry also does not give rise to spin splitting when ${\bf k}\parallel$[001]~\cite{winkler2003spin,Durnev:2022aa}. The vanishing of the spin splitting is due to the $C_{2v}$ symmetry group of the wave vector ${\bf k}$ that parallel to [001]~\cite{winkler2003spin}. This group has only one irreducible double-group representation $\Gamma_5$ which is two-dimensional~\cite{winkler2003spin}. For quasi-one dimensional HgTe nanowires grown along the [001] direction, both the square and the circular cross-sections preserve the $C_{2v}$ symmetry, such that in our case the spin splitting vanishes. While the rectangular cross-section breaks the $C_{2v}$ symmetry to $C_{2}$ symmetry, as studied in Ref.~\cite{PhysRevB.96.115443}, the bulk inversion asymmetry leads to spin splitting.

\section{Discussion and summary}
Here, we briefly discuss the potential influences of the higher-order terms of $k_z$ and $\gamma_3\!-\!\gamma_2$ on the band inversion transition. These higher-order terms may change both the gap-closing-and-reopening position and the anticossing size in the subband structure, but we believe the qualitative behavior of the gap-closing-and-reopening transition will not be influenced. The occurrence of the gap-closing-and-reopening transition in tuning the radius is a consequence of the negative bulk band gap of HgTe (or $\alpha$-Sn), i.e., the negative $\varepsilon_0$ in Eq.~(\ref{eq_Kanemodel}).

The impacts of both the anisotropic term and the bulk inversion asymmetry term of the Kane model on the subband structure of a [001] oriented cylindrical HgTe nanowire are studied in detail. Due to the cylindrical geometry of the nanowire, the bulk inversion asymmetry does not give rise to spin splitting in the subbands. The anisotropic term leads to an anticrossing between the $E_{1}$ and $H_{1}$ subbands at $k_{z}\!=\!0$, and the position of the gap-closing-and-reopening transition  has moved to a $k_{z}\!\neq\!0$ place. The low-energy effective Hamiltonian is a six-band model with two separate $3\!\times\!3$ blocks. The [001] oriented HgTe nanowire is in the topological insulator regime when its radius is larger than 3.45 nm.

%\bibliography{Ref_RL}
%apsrev4-2.bst 2019-01-14 (MD) hand-edited version of apsrev4-1.bst
%Control: key (0)
%Control: author (8) initials jnrlst
%Control: editor formatted (1) identically to author
%Control: production of article title (0) allowed
%Control: page (0) single
%Control: year (1) truncated
%Control: production of eprint (0) enabled
%

\end{document}